\documentclass{osa-article}

\journal{oe}

\usepackage{graphicx}
\usepackage{amsmath}
\usepackage{amsfonts}
\usepackage{amssymb}

\usepackage{upgreek}
\usepackage{caption}

\begin{document}

\title{Towards high power broad-band OPCPA at 3000~nm}

\author{M.~Bridger\authormark{*}, O.~A.~Naranjo-Montoya, A.~Tarasevitch\authormark{**}, and U.~Bovensiepen}

\address{University of Duisburg-Essen, Faculty of Physics, Lotharstrasse 1, 47057 Duisburg, Germany
}
\email{\authormark{*}manuel.bridger@uni-due.de} 
\email{\authormark{**}alexander.tarasevitch@uni-due.de} 



\begin{abstract}
High-energy femtosecond laser pulses in the mid-infrared (MIR) wavelength range
are essential for a wide range of applications from strong-field physics to selectively pump and
probe low energy excitations in condensed matter and molecular vibrations. Here we report
a four stage optical parametric chirped pulse amplifier (OPCPA) which generates ultrashort
pulses at a central wavelength of 3000~nm with 430~$\mu$J energy per pulse at a bandwidth of 490~nm.
Broadband emission of a Ti:sapphire oscillator seeds synchronously the four OPCPA stages at
800~nm and the pump line at 1030~nm. The first stage amplifies the 800~nm pulses in BBO
using a non-collinear configuration. The second stage converts the wavelength to 1560~nm using
difference frequency generation in BBO in a collinear geometry. The third stage amplifies this
idler frequency non-collinearly in KTA. Finally, the fourth stage generates the 3000~nm radiation in
a collinear configuration in LiIO$_3$ due the broad amplification bandwidth this crystal provides.
We compress these pulses to 65~fs by transmission through sapphire. Quantitative calculations
of the individual non-linear processes in all stages verify that our OPCPA architecture operates
close to optimum efficiency at minimum absorption losses, which suggests that this particular
design is very suitable for operation a high average power at multi kHz repetition rates.
\end{abstract}

\section{Introduction}

Development of high power femtosecond laser sources in the MIR range is of high practical interest. On the one hand, this holds for the physics of strong field interactions, because the ponderomotive energy of electrons in the laser field increases quadratically with the laser wavelength. This brings about new possibilities in particle acceleration~\cite{tajima,woodbury}, high order harmonic generation in gases~\cite{popmintchev:pnas,popmintchev:science} and solids~\cite{Hassan:atto}, attosecond electronics~\cite{vampa:nature}, and K$_{\alpha}$ x-ray generation~\cite{kalpha}. On the other hand, femtosecond MIR pulses are essential for time resolved spectroscopy of molecules in gas phase and condensed matter in general. Combined with the above mentioned high harmonic generation MIR pulses also offer a unique possibility of carrying out complimentary pump-probe measurements both in the infrared and soft x-ray spectral range.

Well-established sources of femtosecond pulses in the MIR range are optical parametric oscillators (OPOs) and amplifiers (for the laser MIR sources see~\cite{Cr:ZnSe} and references therein). OPOs synchronously pumped by mode locked Ti:sapphire lasers work at high repetition rates (about 100~MHz) and deliver relatively low energy pulses (below 1~nJ)~\cite{spense96,mcgowan97}. Pulse energies on a microjoule (or even tens of microjoules) level can be achieved with sources utilizing Ti:sapphire or Yb:KGW systems with the repetition rates of about 1~kHz~\cite{petrov97,kaindl00,brida07,vampa:nature,Ishii:19}. Switching from this "short pulse" parametric generation to the OPCPA technique~\cite{dubietis92} allows much higher average power and/or pulse energy using, e.g., picosecond Yb$^{3+}$ or Nd$^{3+}$ pump lasers. In~\cite{mayer}, for example, 42~fs pulses at the wavelength of 3400~nm with the energy of 12~$\mu$J were generated at the repetition rate of 50~kHz(see also~\cite{Kanai:19,Luther:16}). At the repetition rate of 1~kHz energy of 1~mJ at the wavelength of 5~$\mu$m was reached~\cite{Bock:18}. Energies up to 20~mJ at the repetition rate of 20~Hz were demonstrated in~\cite{baltuska:natcom16} (see also~\cite{andriukaitis}). Compared to conventional laser amplifiers high power OPCPA also benefit from much lower thermal load of the amplifying medium.

Several crystals (e.g. LiNbO$_3$, KNbO$_3$, LiIO$_3$, KTA, RTA) support bandwidths needed for sub- 100~fs MIR pulse generation. The largest bandwidths are achieved making use of noncollinear optical parametric amplification (NOPA)~\cite{witte,zaharias}, where the large bandwidth of the signal is achieved by introducing an angular chirp to the idler wave. However, the chirp has to be avoided at the stage, where the MIR radiation is generated as an idler. This makes the choice of a suitable crystals challenging\footnote{In~\cite{Kobayashi98,huang,isaienko} some methods of correction for the idler chirp after NOPA stages have been demonstrated.}. In~\cite{mayer} the broadband amplification (about 400~nm bandwidth with the central wavelength of 3400~nm) was achieved using aperiodically poled MgO:LiNbO$_3$ (see also~\cite{Bigler:18,Thire:18,Zou:19,Rigaud:16}). In~\cite{baltuska:natcom16} the authors make use of a remarkably wide bandwidth of the collinear conversion in KTA type~II interaction at the wavelength of 3900~nm by pumping at the wavelength of about 1000~nm. Unfortunately, KTA crystals start to absorb at the wavelengths longer than 3500~nm~\cite{Hansson:00}, which may lead to limitations in high-repetition-rate systems due to thermal distortions~\cite{tawella:spie,tawella:OL}.

In this paper we report an OPCPA system producing sub-70~fs pulses at the wavelength of 3000~nm with the pulse energy of 430~$\mu$J at a repetition rate of 100~Hz. Our experimental measurements are supported by calculations using a specially developed split-step 2D Matlab code.

\section{Experimental}
\begin{figure}[b]
\centering
\includegraphics[width=12cm]{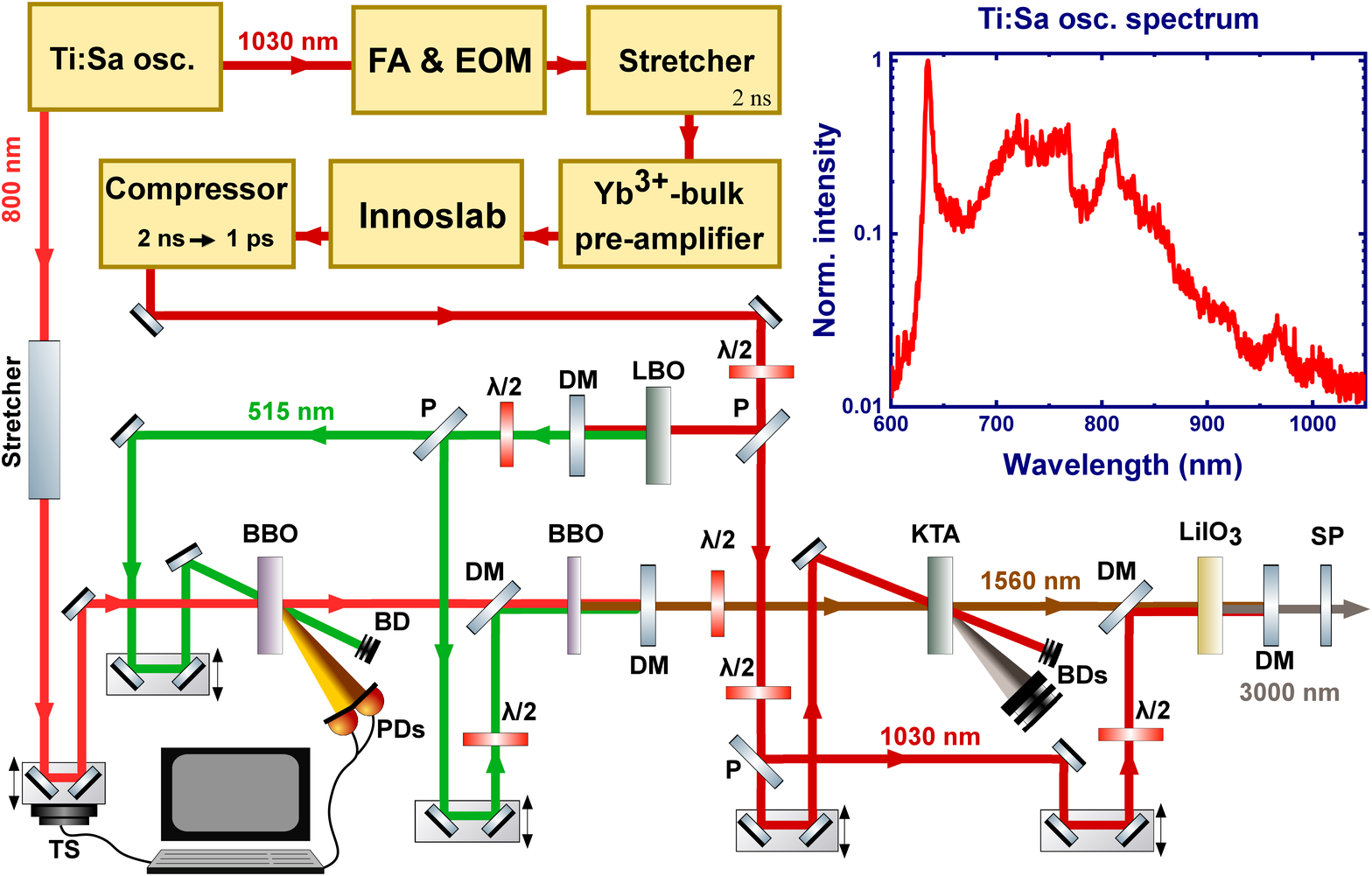}
\caption{OPCPA schematic. A broadband Ti:Sa oscillator is used as a seed source both for the OPCPA itself and for the pump channel. The 6~fs pulses at the wavelength of 800~nm are stretched and used as seed for the four stage OPCPA, consisting of two BBO crystals followed by KTA and LiIO$_3$ stages. The pump channel is seeded with 1030~nm pulses and consists of fiber pre-amplifier with electro-optical modulator (FA \& EOM), stretcher, Yb$^{3+}$ bulk pre-amplifier, Innoslab power amplifier, and compressor. Part of the pump radiation is converted to SH in LBO crystal for pumping of BBO stages. Half-wave plates ($\lambda$/2) are used to control the polarization and together with polarizers (P) to adjust pump intensities. Photodiodes (PDs) together with the computer controlled translation stage (TS) are used to set a proper delay between the pump and the amplified pulses.  DM, BD, and SP stand for dichroic mirrors, beam dumps, and sapphire plate, respectively. The beam diameters are adjusted with the help of telescopes, which are not shown. Inset: spectrum of the Ti:Sa oscillator.}
\label{fig1}
\end{figure}

The OPCPA architecture is depicted in Fig.~\ref{fig1}. The system is seeded by a Ti:sapphire (Ti:Sa) oscillator producing 6~fs pulses at the central wavelength of 800~nm with the energy of 2.5~nJ per pulse and the repetition rate of 80~MHz. The pulse duration was measured directly after the laser output using SPIDER. The spectrum of the Ti:Sa oscillator is presented in the inset to Fig.~\ref{fig1}. A fraction of the emission at the wavelength of 1030$\pm$2~nm is used to seed the OPCPA pump channel.

\subsection{Pump channel}
The above mentioned narrow band Ti:Sa emission at 1030~nm with 30~pJ/pulse passes through an electro optical modulator (EOM), which reduces the pulse repetition rate down to 20~MHz. After a fiber pre-amplifier the pulses with the energy of about 1~nJ are stretched to the duration of about 2~ns. The stretched pulses are amplified in a two stage amplifier with the maximum repetition rate of 300~kHz: a commercial regenerative Yb:KGW amplifier (Amplitude Systems) and subsequently in an Innoslab~\cite{innoslab} power amplifier (Amphos GmbH). The maximum average output power of the pump channel reaches 400~W at repetition rates from about 20 to 300~kHz. In the present work  the repetition rate was 100~Hz . The output pulses with the energy of 20 mJ are compressed to the duration of 1.2~ps. For the stretching and recompression multilayer dielectric gratings (LLNL) with the line density of 1740~1/mm at the angle of incidence of 57$^\circ$ are used. The effective grating distance is 4~m in a folded stretcher and compressor geometry. A 1~mJ fraction of the compressed pulse energy with a flat top spatial profile is frequency doubled in a 2~mm LBO crystal. The crystal is heated to 190$ ^\circ $C in order to achieve non-critical phase matching. The  second harmonic (SH) pulses at the wavelength of 515~nm with the energy of 700~$\mu$J is used to pump the first two OPCPA stages as indicated in Fig.~\ref{fig1}.
\subsection{OPCPA}

The seed pulses provided by the Ti:Sa oscillator are amplified in the first stage in a 3.9~mm thick type~I BBO crystal ($\theta$ = 24.3$^\circ$). The interaction is non-collinearly phase matched with the angle of 2.4$^\circ$ between the pump and the seed beams. Before amplification the 6~fs pulses are stretched with the help of a 10~cm BK7 slab to the duration of 3~ps full width at half maximum (FWHM), which is longer than the pump pulse duration. By changing the delay between the pump and the strongly chirped seed pulses one may choose the central wavelength at which the amplification takes place. The actual delay is set using an active stabilization system~\cite{stabil,stabil:Furch} which utilizes the angular dispersion of an idler, which is intrinsic for non-collinear phase matching. Two parts of the spatially dispersed idler spectrum are measured using two photodiodes. The diode difference signal is used as a process variable for a PID controller, which sets the proper delay using a motorized translation stage.

\begin{figure}[h]
\centering
\includegraphics[width=1.05\textwidth]{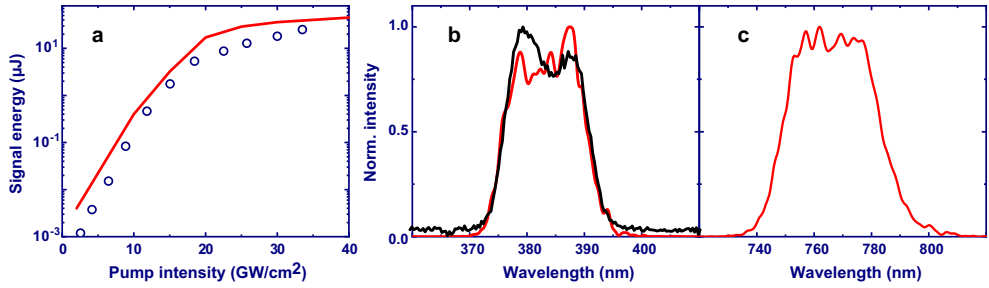}
\caption{Calculated (red line) and measured (open circles) output energy of the first OPCPA stage~(a). Experimental (dashed line) and calculated (red solid line) SH spectra are shown in panel~(b). The calculated spectrum of the signal is shown in panel~(c).}
\label{fig2}
\end{figure}

The experimental and the calculated output energy of the first stage are compared in Fig.~\ref{fig2}a for the amplified spectrum centered around 770~nm. The pump and seed beam diameters are 0.85~mm and 0.64~mm, respectively. The amplified signal pulse energy is measured with a calibrated fast photodiode. It can be seen that at the pump intensity of 35~GW/cm$^2$ the output energy reaches 25~$\mu$J. The saturation of the energy curve starts as the pump intensity approaches 20~GW/cm$^2$. Our 2D calculations show, that above this pump intensity level spatial and temporal distortions of the signal wave rapidly develop. For this reason the stage is subsequently operated at the pump intensity of 20~GW/cm$^2$ which corresponds to the pulse energy of about 100~$\mu$J assuming 0.9~ps pulse duration with the output signal energy of about 10~$\mu$J.

Direct spectral measurements of the amplified pulses faced the challenge of a strong background stemming from unamplified seed pulses at high repetition rate (80~MHz). This background was suppressed by conversion of the signal to the SH using a 60~$\mu$m BBO crystal. The SH spectrum was measured using a grating spectrometer with a CCD (charge coupled device) camera. The SH spectrum together with the calculated one as well as the corresponding calculated signal spectrum are presented in Fig.~\ref{fig2}b and~\ref{fig2}c. Due to good agreement between the two SH spectra in ~\ref{fig2}b, we assume that the real signal spectrum is indeed close to the one drawn in Fig.~\ref{fig2}c. Its bandwidth of about 35~nm FWHM supports 30~fs pulses.
\begin{figure}[t]
\centering
\includegraphics[width=1.05\textwidth]{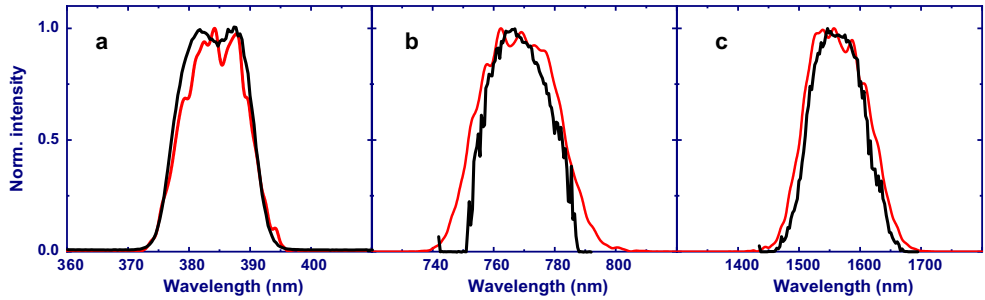}
\caption{Second OPCPA stage: calculated (red solid lines) and measured (dashed line) spectra of the SH of the signal pulse~(a), of the signal~(b) and of the idler~(c), respectively.}
\label{fig3}
\end{figure}

 Now we turn to the second OPCPA stage. A 0.6~mm type~I BBO ($\theta$ = 22.5$^\circ$) crystal is seeded by the signal wave of the first OPCPA stage and is used in the difference frequency generation (DFG). The pump and the signal beams are interacting collinearly in order to avoid the angular chirp of the idler wave. The diameters of both beams are~0.6~mm FWHM. With the pump and the seed pulse energies of 400~$\mu$J and 6~$\mu$J, respectively, the output energy at the central wavelength of 1560~nm is 20~$\mu$J. The corresponding measured and calculated spectra are presented in Fig.~\ref{fig3}. The scanning monochromator coupled with an InGaAs photodiode was used for the spectral measurements. The measured spectral width (FWHM) of the idler of 105~nm supports 43~fs pulses.
\begin{figure}[t]
\centering
\includegraphics[width=1.05\textwidth]{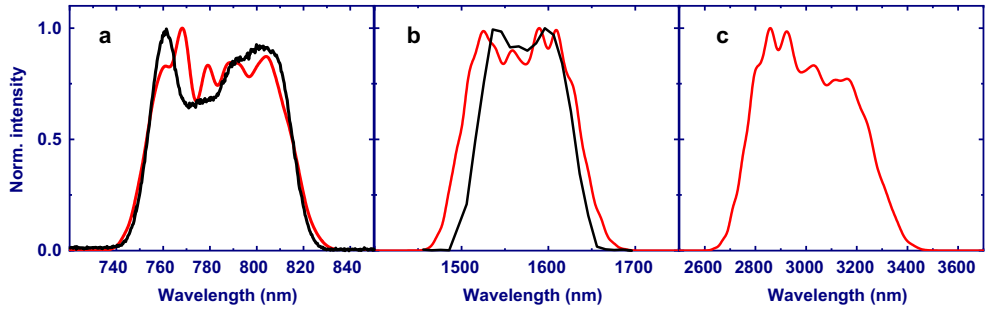}
\caption{Third OPCPA stage: Calculated (red solid lines) and measured (black dashed lines) spectra of the SH of the signal pulse~(a) and of the signal pulse (b). The calculated idler spectrum~(c) is wide enough to support sub 40~fs pulses.}
\label{fig4}
\end{figure}

In the third NOPA OPCPA stage a 3~mm thick type~II KTA crystal with $\theta = 48.6 ^\circ$ is used. The stage is seeded by the idler pulses of the second stage with the pulse energy of 15~$\mu$J at a non-collinearity angle of $ 4.2 ^\circ$. The output energy at the central wavelength of 1560~nm reached 200~$\mu$J with a 140~nm bandwidth FWHM using pump pulses with the energy of 1.5~mJ. The corresponding spectra can be seen in Fig.~\ref{fig4}. The SH and the IR spectra were measured using the spectrometer with the CCD camera and the scanning monochromator with the InGaAs photodiode, respectively.
\begin{figure}[t]
\centering
\includegraphics[width=1.05\textwidth]{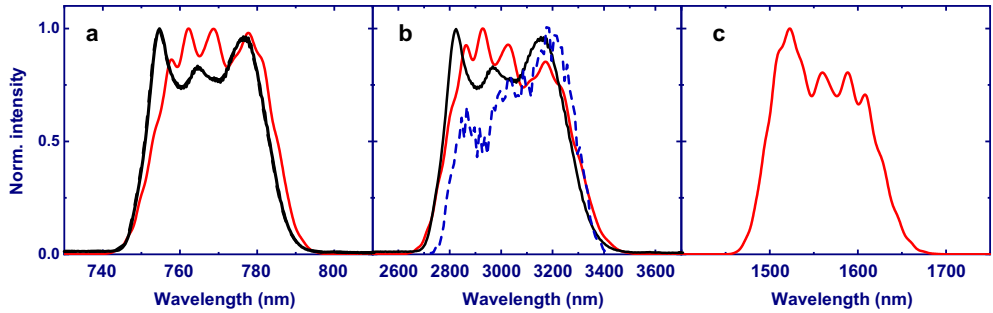}
\caption{Calculated (red solid lines) and measured (black dashed lines) spectra after the fourth stage. Spectrum of the idler pulse up-converted in a thin KTA crystal using pump at 1030~nm (a). Spectra of the idler~(b) and signal~(c).}
\label{fig5}
\end{figure}

\begin{figure}[t]
	\includegraphics[width=1.05\textwidth]{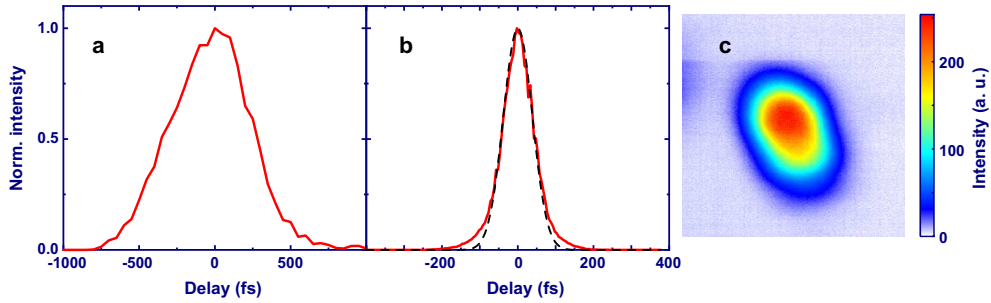}
	\caption{Second order AC trace of uncompressed signal pulses at the wavelength of 1560~nm (a) and of the compressed MIR pulses at the wavelength of 3000~nm (b), red solid curve. Black dashed line in (b) corresponds to a Gaussian function with 90~fs FWHM.  Panel (c) shows the intensity distribution in the 3000~nm beam (recorded using a DataRay camera WinCamD-IR-BB).}
\label{fig6}
\end{figure}

The fourth and last OPCPA stage is used to shift the wavelength to the MIR range. A 4~mm thick type~I LiIO$_3$ crystal, $\theta = 19^\circ$ is pumped with 9~mJ pulses at the wavelength of 1030~nm. With the seed energy of 200~$\mu$J at 1560~nm the output energy of the idler at 3000~nm reaches 430~$\mu$J. The LiIO$_3$ crystal was chosen because its wide amplification bandwidth in collinear geometry. The corresponding spectra are presented in Fig.~\ref{fig5}. They were measured using frequency up conversion in a thin KTA crystal. For the up conversion we used pump pulses at the wavelength of 1030~nm. The spectra of the resulting frequency shifted pulses were centered around 770~nm. Unlike the case of the SH generation, the measured up converted spectra can be directly mapped to the MIR range (Fig.~\ref{fig5}b). This is possible, because the pump spectral width (1.7~nm FWHM) is much smaller and the pulse duration (1.2~ps) longer compared to those of the idler. In order to check the validity of this measurement we have also measured the idler spectrum using a monochromator with a PbSe photodiode (Fig.~\ref{fig5}b, blue dashed line). It can be seen that the directly measured MIR spectrum is cut in the vicinity of 2700-2800 nm. The beam path from the last OPCPA stage to the monochromator is about 3 meters, so this cut can be attributed to the water vapor absorption. The up conversion method delivers an unperturbed spectrum, because the up conversion KTA crystal is placed immediately after the LiIO$_3$ crystal. Another advantage of the up conversion is that the spectra are recorded using a conventional CCD camera practically in a "single shot" mode, which makes the alignment much easier. The idler pulse duration (<500~fs) can be estimated from the autocorrelation measurement of the signal pulse (see Fig.~\ref{fig6}a).

According to Fig.~\ref{fig5}b the bandwidth of the 3000~nm pulses exceeds 490~nm FWHM. This bandwidth supports 35~fs pulse duration, if properly compressed. In this work we have used a simple compression using sapphire plates to compensate only for the second order dispersion. A second order autocorrelation (AC) function using a 0.5~cm sapphire plate is presented in Fig.~\ref{fig6}b. From the measured AC width of 90~fs (FWHM) we estimate the pulse duration to be about 65~fs using the deconvolution factor of 1.4 for Gaussian pulses. In all the stages the calculated B-integral at maximum intensity was below $\pi$/4, assuming the nonlinear refraction index $\tilde{n}_{2}$ to be 8$\times$10$^{-16}$, 17$\times$10$^{-16}$, and $\leq$20$\times$10$^{-16}$~cm$^2$/W for BBO, KTA, and LiIO$_3$, respectively. The intensity distribution in the 3000~nm beam taken at 1~m distance from the LiIO$_3$ crystal with the help of an infrared CCD camera is shown in Fig.~\ref{fig6}c.

\section{Conclusions and outlook}
In summary, we report an OPCPA system generating $430~\mu$J pulses at the wavelength of 3000~nm with the bandwidth of 490~nm FWHM and the repetition rate of 100 Hz. Using a simple bulk compressor the pulses were compressed to the duration of about 65~fs. The reported pulse bandwidth supports 35~fs pulses, which can be probably achieved by compensation of higher order dispersion.

The combination of BBO, KTA, and LiIO$_3$ crystals in the four OPCPA stages allows keeping a broad bandwidth at the center wavelength of 3000~nm with little absorption in the crystals. KTA and LiIO$_3$ crystals are available with relatively low absorption at the pump wavelength ($\alpha_{KTA}\approx$ 0.01~\%/cm~\cite{tawella:spie}) and $\alpha_{LiIO_3}$< 0.02~\%/cm~\cite{LiIO3:absorption}). This may allow MIR pulse production with much higher repetition rates and average powers, e.g. average power >5~W is achievable at the repetition rate of 20~kHz keeping the same pulse energy.

\section{Funding}
We acknowledge financial support by the Deutsche Forschungsgemeinschaft through SFB 1242 (project number 278162697, TP A05).

\section{Acknowledgments}
We thank Frank Meyer for his experimental support and initial development of the Matlab code. We are also very grateful to A. Baltu\v{s}ka, A. Pug\v{z}lys, and S. Ali\v{s}auskas for their valuable help in early stages of the work.


\end{document}